\documentclass{article}

\PassOptionsToPackage{numbers, compress}{natbib}

\usepackage[final]{neurips_2019}
\usepackage[utf8]{inputenc} 
\usepackage[T1]{fontenc}    
\usepackage{hyperref}       
\usepackage{url}            
\usepackage{booktabs}       
\usepackage{amsfonts}       
\usepackage{nicefrac}       
\usepackage{microtype}      
\usepackage[pdftex]{graphicx} 
\usepackage[font=small,labelfont=bf,skip=5pt]{caption}

\hypersetup{hidelinks}

\title{Neural Ordinary Differential Equations for \\Semantic Segmentation of Individual Colon Glands}

\author{%
  Hans Pinckaers \\
  Computational Pathology Group\\
  Radboud University Medical Center\\
  The Netherlands \\
  \texttt{hans.pinckaers@radboudumc.nl} \\
  \And
  Geert Litjens \\
  Computational Pathology Group\\
  Radboud University Medical Center\\
  The Netherlands \\
  \texttt{geert.litjens@radboudumc.nl} \\
}

\begin{document}

\maketitle

\begin{abstract} 
Automated medical image segmentation plays a key role in quantitative research and diagnostics. Convolutional neural networks based on the U-Net architecture are the state-of-the-art. A key disadvantage is the hard-coding of the receptive field size, which requires architecture optimization for each segmentation task. Furthermore, increasing the receptive field results in an increasing number of weights. Recently, Neural Ordinary Differential Equations (NODE) have been proposed, a new type of continuous depth deep neural network. This framework allows for a dynamic receptive field at a constant memory cost and a smaller amount of parameters. We show on a colon gland segmentation dataset (GlaS) that these NODEs can be used within the U-Net framework to improve segmentation results while reducing memory load and parameter counts.  
\end{abstract}

\section{Introduction}

Automated medical image segmentation plays a key role in quantitative research\cite{Sirinukunwattana2017, Litjens2014} and diagnostics\cite{Litjens2017}. The performance of semantic segmentation networks depends partly on the receptive field of those networks. Wider receptive views allow for increased use of context and often comes at the benefit of higher accuracy\cite{chen2018encoder, Brugger2019}, but a limited amount of GPU memory forces a trade-off between network depth, width, batch size, and input image size.

The current de facto standard network architecture for segmentation in medical images are convolutional neural networks, specifically those following the U-Net architecture\cite{ronneberger2015u}. The original U-Net has a receptive field of 187 pixels at the computational cost of 30 million parameters. This receptive field of U-Net is static and bounded by the number of layers (23 convolutional layers) and downsample operations. Adding layers or whole levels can increase the receptive field, however, this comes at the cost of more parameters, computation and memory requirements. Furthermore, receptive field sizes need to be optimized for every individual segmentation task.

We propose to use a new family of neural networks, called Neural Ordinary Differential Equations (NODE)\cite{Chen2018}, a continuous depth deep neural network, within the U-Net framework to memory-efficiently provide an adaptive receptive view. We are the first to apply this technique on a segmentation task and show a proof-of-concept on the GlaS challenge dataset\cite{Sirinukunwattana2017}. This public challenge involved segmenting individual colon glands in histopathology images. We show that these NODEs can be used within the U-Net framework to improve segmentation results.

\textbf{Related work} \quad To increase the receptive field of a network, several extensions to the U-Net architecture have been proposed including dilated convolutions\cite{Folle2019, Li2019, Devalla2018, chen2018encoder} and reversible blocks\cite{Brugger2019}. Our approach is similar to reversible blocks, with the additional benefits of an adaptive receptive field per task and image.
\pagebreak

\textbf{Neural ODEs} \quad We will briefly introduce NODEs, for a more extensive write up we refer to the paper by Chen et al., 2018\cite{Chen2018}. NODEs can be understood as a continuous depth equivalent to residual neural networks (ResNets\cite{He2016}). Every block with parameters $\theta$ of a residual neural network calculates some transformation $f(h_t)$ on its input $h_t$:
\begin{equation}
  h_{t+1} = h_t + f(h_t, \theta_t)
\end{equation}
where $t \in \{0 ... T\}$, $h_t \in \mathbb{R}^d$, and $f$ a differentiable function. In ResNets, $f$ consists of several convolutional layers. The update with residual $f(h_t)$ can be seen as a $\Delta t=1$ step of an Euler discretization of a continuous transformation. When we let $\Delta t \to 0$ we take more, smaller, steps using more layers, which in the limit becomes an ordinary differential equation (ODE), specified by a neural network:
\begin{equation}
  \lim_{\Delta t \to 0} = \frac{h_{t+\Delta t}-h_t}{\Delta t} = \frac{\delta h_t}{\delta t} = f(h(t), t, \theta)
\end{equation}
ODEs can be solved using standard ODE solvers such as Runge-Kutta\cite{Runge1895,Kutta1901a}. To update the weights of the convolutional layers, we would need to backpropagate through the solver. This can be done in the same way as a regular CNN, however, this is not memory-efficient. Specifically, an ODE solver might need hundreds of function evaluations, leading to exploding memory requirements. Instead, the ODE solver is regarded as a `black box solver' and the gradients are computed via the \textit{adjoint method}. This approach involves another ODE that goes backward in time starting with the gradients of the original output w.r.t. the loss. Gradients w.r.t. the parameters $\theta$ are calculated by automatic differentiation, which can efficiently be performed during the reverse-mode second ODE (See \textbf{Algorithm 1} in Chen et al., 2018\cite{Chen2018}).

A NODE network has several advantages for semantic segmentation. (1) They are memory efficient since intermediate computations (e.g. activation maps) do not need to be stored. (2) They provide an adaptive receptive view, both during training and inference, since modern ODE solvers can alter the number of function evaluations (e.g. the number of times the convolutional layers are applied) to minimize approximation error. This also allows the end-user to make trade-offs between accuracy and inference speed at test time, to fit hardware requirements of embedded systems, for example. (3) Sharing parameters across the sequential layers (function evaluations) reduces the number of parameters and thus prevents overfitting.

\begin{figure*} 
    \centering
    {\includegraphics[width=\textwidth]{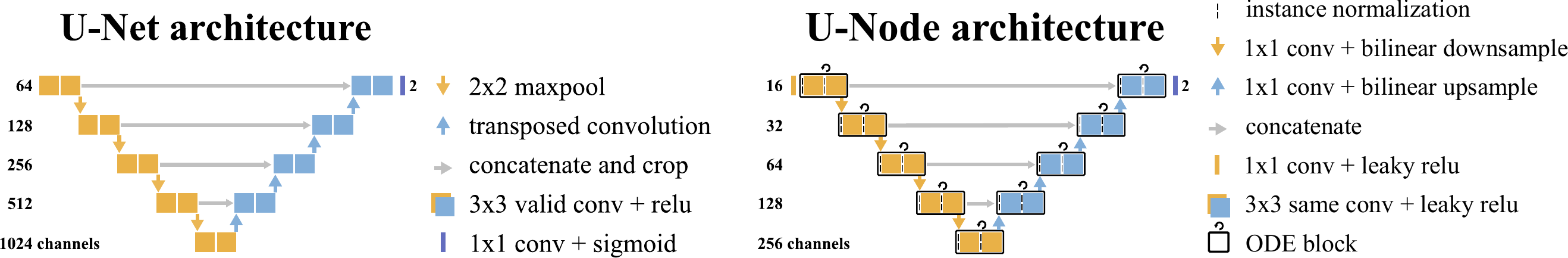}}%
    \caption{Schematic overview of U-Net and the proposed U-Node network. The U-ResNet architecture is equivalent to U-Node except for the ODE blocks.}
    \label{figure:networks}
\end{figure*}

\section{Experiments}
\begin{figure}[h] 
    \centering
    {\includegraphics[width=\textwidth]{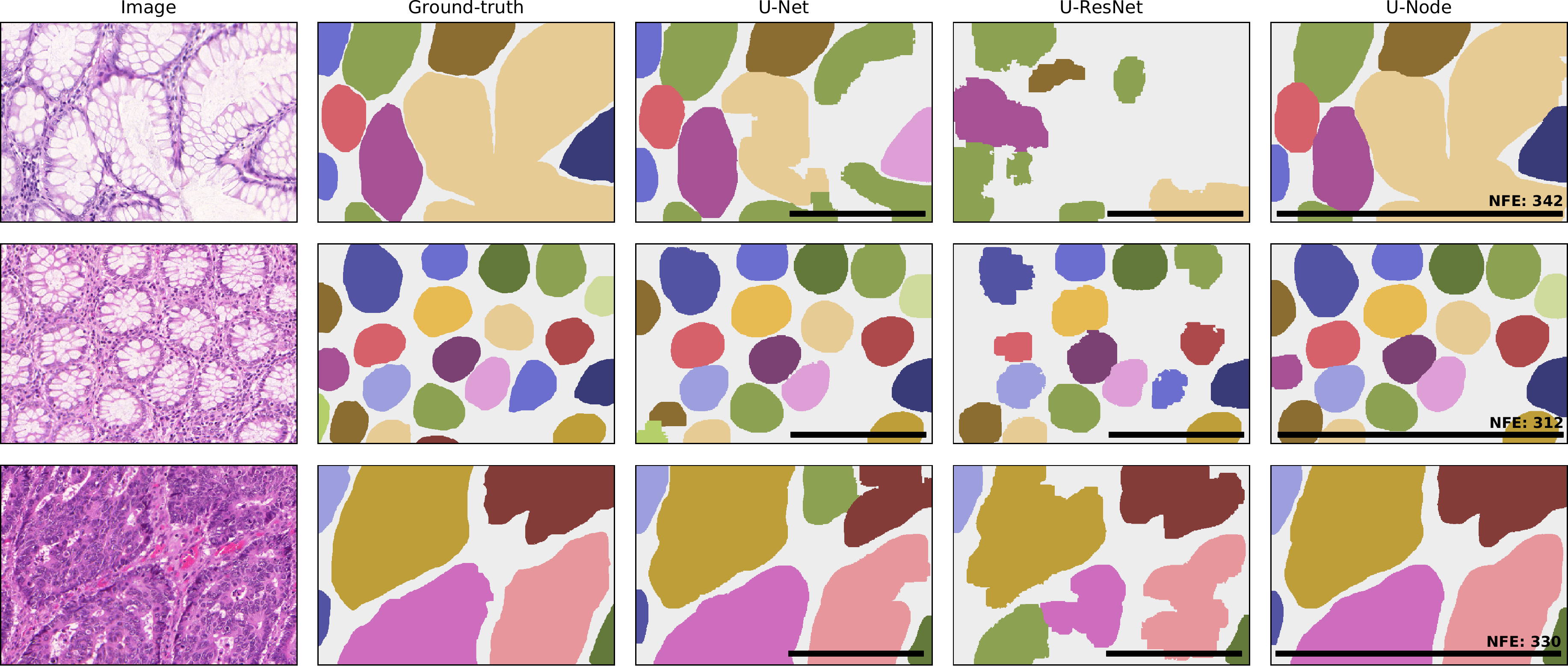}}%
    \caption{Example patches of the test set with clear differences between the models. Each color denotes a gland. The black bar denotes the receptive field. NFE denotes the number of function evaluations in the network. Difficult images (first and last row) seem to require more evaluations.}
\label{figure:results}
\end{figure}

The GlaS dataset\cite{Sirinukunwattana2017} consists of a training set with 85 images and a test set of 80 images. The majority are $775 \times 522$ pixel patches from scanned whole-slide histology images of the colon, where epithelial glands have been annotated (Fig. \ref{figure:results}, column 1 and 2). The test dataset is divided into two subsets; subset A (60 images) released earlier and subset B (20 images) released during the original MICCAI workshop. We report results on the combined test set and the individual subsets.

We train three models (see Fig. \ref{figure:networks}): (1) A baseline U-Net model\cite{ronneberger2015u}, with 30m parameters; (2) A U-Net with less filters and ODE blocks, termed U-Node, with 2m parameters; (3) The U-Node network, but conventionally trained, equal to one function evaluation per `ODE' block, with 2m parameters, termed U-ResNet.\footnote{Source code can be found at \url{https://github.com/DIAGNijmegen/neural-odes-segmentation}} 

We train the models to predict the full segmentation mask and an eroded version to separate individual glands when post-processing. To handle the different image sizes and allow one image to fit on the GPU we downscale $1.5\times$ and reflection pad to $352 \times 512$ px. At train time we apply the following random augmentations: translation, flipping, rotation, elastic transformation, and color jitter. We use the Adam optimizer\cite{kingma:adam}, with mini-batches of eight images, a learning rate of $10^{-3}$ ($10^{-4}$ for U-Net otherwise training was unstable), and cross-entropy loss. We used ODE solvers from the torchdiffeq python package\cite{Chen2018} and used the fifth-order ``dopri5'' solver, with a $10^{-3}$ tolerance. We randomly take ten images from the training set as a tuning set. We trained for 600 epochs. We did not use early stopping, as the validation loss plateaued. At test time, we apply test-time augmentation and average predictions over the original, horizontal, and vertical flipped image.

\begin{table}[h]
\footnotesize
\centering
\caption{GlaS challenge metrics for the total test set and subsets (A, B)}
\label{tab:glas}
\begin{tabular}{lllll}
\toprule
\textbf{Method} & \textbf{Object Dice (A, B)} & \textbf{F1 score (A, B)} & \textbf{Hausdorff* (A, B)} & \textbf{Notes} \\
\midrule
U-Net & 0.868 (0.884, 0.819) & 0.841 (0.865, 0.768) & 69.6 (55.6, 111) & 30m parameters\\  
U-ResNet & 0.757 (0.789, 0.660) & 0.689 (0.743, 0.523) & 122 (97.3, 199) & 2m parameters\\  
U-Node & \textbf{0.881} (0.893, 0.842) & \textbf{0.861} (0.882, 0.801) & \textbf{59.5} (48.6, 92.4) & 2m parameters\\
\midrule
\midrule
Chen et al.\cite{Chen2016} & 0.868 (0.897, 0.781) & 0.863 (0.912, 0.716) & 74.2 (45.4, 160.3) & GlaS winner\\
Graham et al.\cite{Graham2019} & \underline{\textbf{0.902}} (0.919, 0.849) & \underline{\textbf{0.896}} (0.920, 0.824) & \underline{\textbf{54.7}} (41.0, 95.7) & SOTA\\
\bottomrule
\multicolumn{4}{l}{*a lower Hausdorff distance is better.} \\
\end{tabular}
\end{table}

\section{Results}
We evaluate our models using the challenge metrics\cite{Sirinukunwattana2017}; a weighted Dice coefficient per gland (object Dice), a gland detection F1 score, and a weighted shape similarity metric based on the Hausdorff distance (where lower is better). Table \ref{tab:glas} shows the performance of our models.

\section{Discussion and conclusion}
This study investigated whether incorporating neural ordinary differential equations could be beneficial in semantic segmentation. The results show that the proposed U-Node network can efficiently use less parameters and improve segmentation compared to U-Net and U-ResNet. Qualitative result in Fig. \ref{figure:results} show that overall U-Node produces segmentations with less noise. Furthermore, the adaptive receptive field seems to help segment larger glands. 

Due to the increased number of convolutional operations compared to U-Net, training is computationally heavier and slower for U-Node. Training with lower tolerances can provide a speed up, but at the cost of performance. We used the same number of levels as U-Net to force compression of the latent space, however, the training time of the model can be improved by reducing the number of ODEs, for example by only using ODEs in the encoder (down-path), or by decreasing the number of levels.

Graham et al.\cite{Graham2019} train a rotation equivariant network using group equivariant convolutions\cite{Cohen2016} to reach state-of-the-art performance on the GlaS challenge. A logical next step would be to combine our U-Node model with group equivariant convolutions, providing memory-efficient group equivariance for rotations with a large adaptive receptive field. 

\subsubsection*{Acknowledgments}

We would like to thank Jasper Linmans for the useful discussions.

\medskip

\bibliographystyle{utphys}
\bibliography{library}

\providecommand{\href}[2]{#2}\begingroup\raggedright\begin{thebibliography}{10}

\bibitem{Sirinukunwattana2017}
K.~Sirinukunwattana, J.~P. Pluim, H.~Chen, X.~Qi, P.-A. Heng, Y.~B. Guo, L.~Y.
  Wang, B.~J. Matuszewski, E.~Bruni, U.~Sanchez, A.~B{\"{o}}hm, O.~Ronneberger,
  B.~B. Cheikh, D.~Racoceanu, P.~Kainz, M.~Pfeiffer, M.~Urschler, D.~R. Snead,
  and N.~M. Rajpoot, ``{Gland segmentation in colon histology images: The glas
  challenge contest},''
  \href{http://dx.doi.org/10.1016/j.media.2016.08.008}{{\em Medical Image
  Analysis} {\bfseries 35} 2017, 489--502}.

\bibitem{Litjens2014}
G.~Litjens, R.~Toth, W.~van~de Ven, C.~Hoeks, S.~Kerkstra, B.~van Ginneken,
  G.~Vincent, G.~Guillard, N.~Birbeck, J.~Zhang, R.~Strand, F.~Malmberg, Y.~Ou,
  C.~Davatzikos, M.~Kirschner, F.~Jung, J.~Yuan, W.~Qiu, Q.~Gao, P.~E. Edwards,
  B.~Maan, F.~van~der Heijden, S.~Ghose, J.~Mitra, J.~Dowling, D.~Barratt,
  H.~Huisman, and A.~Madabhushi, ``{Evaluation of prostate segmentation
  algorithms for MRI: The PROMISE12 challenge},''
  \href{http://dx.doi.org/10.1016/j.media.2013.12.002}{{\em Medical Image
  Analysis} {\bfseries 18} no.~2, 2014, 359--373}.

\bibitem{Litjens2017}
G.~Litjens, T.~Kooi, B.~E. Bejnordi, A.~A.~A. Setio, F.~Ciompi, M.~Ghafoorian,
  J.~A. van~der Laak, B.~van Ginneken, and C.~I. S{\'{a}}nchez, ``{A survey on
  deep learning in medical image analysis},'' 2017.

\bibitem{chen2018encoder}
L.-C. Chen, Y.~Zhu, G.~Papandreou, F.~Schroff, and H.~Adam, ``{Encoder-decoder
  with atrous separable convolution for semantic image segmentation},'' in {\em
  Proceedings of the European conference on computer vision (ECCV)},
  pp.~801--818.
\newblock 2018.

\bibitem{Brugger2019}
R.~Br{\"{u}}gger, C.~F. Baumgartner, and E.~Konukoglu, ``{A Partially
  Reversible U-Net for Memory-Efficient Volumetric Image Segmentation},''
  \href{http://arxiv.org/abs/1906.06148}{{\ttfamily arXiv:1906.06148}}.

\bibitem{ronneberger2015u}
O.~Ronneberger, P.~Fischer, and T.~Brox, ``{U-net: Convolutional networks for
  biomedical image segmentation},'' in {\em International Conference on Medical
  image computing and computer-assisted intervention}, pp.~234--241, Springer.
\newblock 2015.

\bibitem{Chen2018}
T.~Q. Chen, Y.~Rubanova, J.~Bettencourt, and D.~Duvenaud, ``{Neural Ordinary
  Differential Equations},'' \href{http://arxiv.org/abs/1806.07366}{{\ttfamily
  arXiv:1806.07366}}.

\bibitem{Folle2019}
L.~Folle, S.~Vesal, N.~Ravikumar, and A.~Maier,
  \href{http://dx.doi.org/10.1007/978-3-658-25326-4_18}{``{Dilated deeply
  supervised networks for hippocampus segmentation in MRI},''} in {\em
  Informatik aktuell}, pp.~68--73.
\newblock 2019.
\newblock \href{http://arxiv.org/abs/1903.09097}{{\ttfamily arXiv:1903.09097}}.

\bibitem{Li2019}
H.~Li, A.~Zhygallo, and B.~Menze,
  \href{http://dx.doi.org/10.1007/978-3-030-11723-8_39}{``{Automatic brain
  structures segmentation using deep residual dilated U-Net},''} in {\em
  Lecture Notes in Computer Science (including subseries Lecture Notes in
  Artificial Intelligence and Lecture Notes in Bioinformatics)}, vol.~11383
  LNCS, pp.~385--393.
\newblock 2019.
\newblock \href{http://arxiv.org/abs/1811.04312}{{\ttfamily arXiv:1811.04312}}.

\bibitem{Devalla2018}
S.~K. Devalla, P.~K. Renukanand, B.~K. Sreedhar, S.~Perera, J.-M. Mari, K.~S.
  Chin, T.~A. Tun, N.~G. Strouthidis, T.~Aung, A.~H. Thiery, and M.~J.~A.
  Girard, ``{DRUNET: A Dilated-Residual U-Net Deep Learning Network to
  Digitally Stain Optic Nerve Head Tissues in Optical Coherence Tomography
  Images},'' \href{http://dx.doi.org/10.1167/iovs.17-22617}{{\em Invest.
  Ophthalmol. Vis. Sci.} {\bfseries 59} no.~1, 2018, 63--74},
  \href{http://arxiv.org/abs/1803.00232}{{\ttfamily arXiv:1803.00232}}.

\bibitem{He2016}
K.~He, X.~Zhang, S.~Ren, and J.~Sun,
  \href{http://dx.doi.org/10.1109/CVPR.2016.90}{``{Deep residual learning for
  image recognition},''} in {\em Proceedings of the IEEE Computer Society
  Conference on Computer Vision and Pattern Recognition}, vol.~2016,
  pp.~770--778.
\newblock IEEE Computer Society, 2016.

\bibitem{Runge1895}
C.~Runge, ``{Ueber die numerische Aufl{\"{o}}sung von
  Differentialgleichungen.},'' {\em Mathematische Annalen} {\bfseries 46} 1895,
  167--178.

\bibitem{Kutta1901a}
W.~Kutta, ``{Beitrag zur n{\"{a}}herungsweisen Integration totaler
  Differentialgleichungen},'' {\em Zeitschrift f{\"{u}}r Mathematik und Physik}
  {\bfseries 46} 1901, 435--453.

\bibitem{kingma:adam}
D.~P. Kingma and J.~Ba, ``{Adam: A method for stochastic optimization},'' in
  {\em International Conference on Learning Representations (ICLR)}.
\newblock 2015.

\bibitem{Chen2016}
H.~Chen, X.~Qi, L.~Yu, and P.~A. Heng,
  \href{http://dx.doi.org/10.1109/CVPR.2016.273}{``{DCAN: Deep Contour-Aware
  Networks for Accurate Gland Segmentation},''} in {\em Proceedings of the IEEE
  Computer Society Conference on Computer Vision and Pattern Recognition},
  vol.~2016-Decem, pp.~2487--2496.
\newblock 2016.
\newblock \href{http://arxiv.org/abs/1604.02677}{{\ttfamily arXiv:1604.02677}}.

\bibitem{Graham2019}
S.~Graham, D.~Epstein, and N.~Rajpoot, ``{Rota-Net: Rotation Equivariant
  Network for Simultaneous Gland and Lumen Segmentation in Colon Histology
  Images},'' in {\em Digital Pathology}, C.~C. Reyes-Aldasoro, A.~Janowczyk,
  M.~Veta, P.~Bankhead, and K.~Sirinukunwattana, eds., pp.~109--116.
\newblock Springer International Publishing, Cham, 2019.

\bibitem{Cohen2016}
T.~S. Cohen and M.~Welling, ``{Group equivariant convolutional networks},'' in
  {\em 33rd International Conference on Machine Learning, ICML 2016}, vol.~6,
  pp.~4375--4386.
\newblock 2016.
\newblock \href{http://arxiv.org/abs/1602.07576}{{\ttfamily arXiv:1602.07576}}.

\end{thebibliography}\endgroup

\end{document}